\documentclass[prl,twocolumn,letterpaper,amsmath,showpacs,floatfix,longbibliography]{revtex4-2}
\RequirePackage{graphicx}
\usepackage{placeins}
\usepackage{hyperref}
\newcommand{\D}{\mathrm d}
\newcommand{\E}{\mathrm e}
\newcommand{\I}{\mathrm i}
\renewcommand{\vec }{\mathbf }
\newcommand{\pD}{\partial }
\newcommand{\curl}{\mathrm {curl}}
\newcommand{\diver}{\mathrm {div}}

\begin{document}

\title{Inverse cascade of gravity waves in the presence of condensate: numerical simulation.}

\date{\today}

\author{Alexander\,O.~Korotkevich}
\email{alexkor@math.unm.edu}
\affiliation{Department of Mathematics and Statistics, University of New Mexico, MSC01 1115, 1 University of New Mexico, Albuquerque, NM 87131-0001, USA}
\affiliation{L.\,D.~Landau Institute for Theoretical Physics RAS, Prosp. Akademika Semenova 1A, Chernogolovka, Moscow region, 142432, Russian Federation}
\pacs{47.27.ek, 47.35.-i, 47.35.Jk}

\begin{abstract}
During the set of direct numerical simulations of the forced isotropic turbulence of surface gravity waves in the framework of primordial dynamical equations, the universal inverse cascade spectrum was observed. The slope of the spectrum is the same (in the margin of error) for different levels of pumping and nonlinearity as well as dissipation present in the system. In all simulation runs formation of the inverse cascade spectrum was accompanied by appearance of strong long wave background (condensate). The observed slope of the spectrum $\sim k^{-3.07}$ is different from the constant wave action flux solution predicted by the Theory of Wave Turbulence $\sim k^{-23/6}$.
\end{abstract}

\maketitle
\FloatBarrier
\section{Introduction}
The Waves Turbulence Theory (WTT) (see e.g.~\cite{ZLF1992,Nazarenko2011}) describes evolution of a distribution function for weakly nonlinear waves. One of the most important applications of WTT is a problem of wave forecasting: statistical description of evolution of the wave field in a sea or an ocean. Most of the current operational waves forecasting models are based either directly on Hasselmann waves kinetic equation~\cite{Hasselmann1962} (WKE) for surface gravity waves (when one neglects capillary effects) with extra phenomenological terms or on its variations~\cite{KCDHHJ1994,CavaleriEtAl2007}. Thus, verification of the applicability of the WKE for different setups is an important practical question. Two constant flux solutions of WKE for gravity waves, corresponding to direct~\cite{ZF1967} and inverse~\cite{ZZ1982} cascades, were found by Zakharov and co-authors~\cite{Zakharov1992,ZLF1992,Nazarenko2011}. These Kolmogorov-Zakharov (KZ) solutions are formulated for inertial intervals: ranges of scales where dynamics is determined by nonlinear interaction of waves and direct influence of dissipation or pumping is negligible. If one could confirm observation of these solutions in a field, laboratory or numerical experiment, this would be a strong argument in support of applicability of WKE in particular conditions.

While the spectrum corresponding to the direct cascade of energy to small scales was observed both experimentally~\cite{Donelan1985,Hwang2000} and numerically~\cite{Onorato2002,DKZ2003grav,DKZ2004,Yokoyama2004,LNP2006}, the spectrum of inverse cascade of wave action from smaller to larger scales appeared to be way harder case. Although frequency downshift, which can be explained by inverse cascade, was observed in direct numerical simulations (DNSs) of so called decaying turbulence (dynamics of initial conditions spectrum without energy being pumped into the system by external forcing)~\cite{Onorato2002,ZKPD2005,ZKPR2007,KPRZ2008}, it could not be a substitute for inverse cascade spectrum, which can be obtained only in the case of forced waves turbulence. Perhaps, the first attempt toward this goal was article~\cite{AS2006PRL}, where initial stage of formation of KZ-spectrum was demonstrated, but the range of scales with power-like spectrum was definitely not sufficient to determine the slope. The attempt to observe both inverse and direct cascades at the same time~\cite{Korotkevich2008PRL} has resulted in observation of formation of inverse cascade and a {\it condensate} (strong long wave background), which affected even direct cascade spectrum, but range of scales in the inverse cascade region was again insufficient to determine the slope accurately enough. Description of some of the laboratories experiments, where observation of inverse cascade was attempted, can be found in~\cite{NL2016}. The major problem in early laboratory experiments~\cite{DLF2011} was small size of the basin which required to excite waves in the capillary-gravity crossover region~\cite{FM2022}. Even in later experiments in a relatively large wave tank~\cite{FalconEtAl2020}, both the dynamical range and finite size effects prevented clear observation of the slope of the inverse cascade. It should be mentioned that in all numerical and at least some of the laboratory experiments mentioned above formation of the condensate was observed (although, not always noted).

Importance of condensate influence upon the gravity waves spectrum was demonstrated in~\cite{Korotkevich2008PRL} and investigated in more details in~\cite{Korotkevich2012MCS}. At least one of the important mechanisms of the condensate formation is prevention of the nonlinear interactions due to increasingly more important discreetness of the homogeneous wavenumbers grid, typical for both DNS with periodic boundary conditions and relatively small laboratory basins, like in~\cite{DLF2011}, which means that condensate is not purely numerical artefact. Importance of discreetness of wavenumbers grid for nonlinear interactions was noted a long time ago for waves in resonators~\cite{Kartashova1991,Kartashova2012} and investigated in details with direct application to WTT~\cite{DKZ2003cap,Nazarenko2006,LNP2006,KDZ2016}. Although, there are exact resonances present on a discreet homogeneous grid of wavevectors~\cite{ZP2022}, the quasi-resonances due to nonlinear broadening of the resonance curve play very important role for turbulent fluxes in simulations of WTT~\cite{AS2006JFM}. As the inverse cascade spectrum propagates further from the pumping region to the smaller wavenumbers (larger scales), eventually nonlinear interaction through quasi-resonances is arrested by the discreetness and the flux only brings wave action without further propagation, resulting in accumulation of it at some large scale. This results in formation of a strong (at least order of magnitude larger than even closest harmonics) long wave background, which we call condensate. The power-like inverse cascade spectrum can be observed in the inertial interval between condensate and pumping regions, like in laboratory experiments~\cite{DLF2011,NL2016} or in DNS~\cite{AS2006PRL,Korotkevich2008PRL}. In these simulations the inertial interval was too short to allow to determine the slope of inverse cascade with reasonable accuracy and compare it with WTT prediction. Taking into account extremely slow formation of inverse cascade (meaning smaller resolution for faster computations) and at the same time demand of a reasonable dynamic range for determining the slope of the spectrum, one needs to find a compromise between these contradicting requirements. 

In this Letter we present results of DNSs in the framework of primordial dynamical equations of the forced turbulence of surface gravity waves and formation of inverse cascade with a power-like spectrum. Simulations were performed for different levels of pumping, resulting in different nonlinearity levels, different parameters of dissipation, for a long ($\sim 10^6$ periods of central pumping harmonic) period of time. In all the cases we obtained condensate formation and were able to determine the (universal) slope of power-like spectrum in the inertial interval, virtually the same for all simulations, which is different from the inverse cascade spectrum predicted by WTT. Deviation of the new universal spectrum from the WTT inverse cascade constant flux spectrum might be due to interaction with the condensate, which, among other things, changes significantly dispersion relation of the waves~\cite{Korotkevich2013JETPL}.

\section{Problem formulation.}
We consider a potential flow (velocity of the fluid is $\vec v = \vec\nabla\Phi$) of an ideal incompressible fluid of infinite depth.
The elevation of the 2D-surface over 3D-fluid from the steady state is described by a function $\eta(\vec r;t)$,
where $\vec r=(x,y)^{T}$ is the coordinate vector in a horizontal plane. Velocity potential on the surface is $\psi(\vec r;t)=\Phi|_{z=\eta(\vec r; t)}$.
The system is Hamiltonian~\cite{Zakharov1968} with respect to variables $\eta$ and $\psi$. An average slope of the surface $\mu=\sqrt{\langle|\vec\nabla\eta(\vec r)|^2\rangle}$, also called steepness, in most of the observations is a small parameter $\mu \ll 1$.
One can expand the Hamiltonian in powers of $\mu$ (see detailed derivation in Supplemental Materials) and obtain Hamiltonian equations:
\begin{equation}
\label{eta_psi_system}
\begin{array}{lcl}
\displaystyle
\dot \eta &=& \hat k  \psi - (\nabla (\eta \nabla \psi)) - \hat k  [\eta \hat k  \psi] +\\
\displaystyle
		&&+ \hat k (\eta \hat k  [\eta \hat k  \psi]) + \frac{1}{2} \Delta [\eta^2 \hat k \psi] + 
		\frac{1}{2} \hat k [\eta^2 \Delta\psi] - F^{-1}[\gamma_k \eta_{\vec k}],\\
\displaystyle
\dot \psi &=& - g\eta - \frac{1}{2}\left[ (\nabla \psi)^2 - (\hat k \psi)^2 \right] - \\
\displaystyle
		&& - [\hat k  \psi] \hat k  [\eta \hat k  \psi] - [\eta \hat k  \psi]\Delta\psi - F^{-1}[\gamma_k \psi_{\vec k}]
		+ F^{-1}[f_k].
\end{array}
\end{equation}
Artificial pumping and damping terms will be described in details later in~\eqref{pumping_damping}. Here $\hat k$ is a linear integral operator 
$\hat k =\sqrt{-\Delta}$, such that $\hat k f_{\vec r}$ in $k$-space corresponds to multiplication of Fourier (in horizontal $XY$-plane) coefficients $f_{\vec k}$:
\begin{equation*}
\label{FourierTrans}
\begin{array}{cl}
\displaystyle
\hat F [f_{\vec r}] &= f_{\vec k} = \frac{1}{L_x L_y} \int\limits_{0}^{L_x}\int\limits_{0}^{L_y} f_{\vec r} e^{-\I {\vec k} {\vec r}} \D\vec r,\\
\displaystyle
\hat F^{-1} [f_{\vec k}] &= f_{\vec r} = \sum\limits_{\vec k} f_{\vec k} e^{\I {\vec k} {\vec r}}.
\end{array}
\end{equation*}
by $k=|\vec k|=\sqrt{k_{x}^2 + k_{y}^2}$. For gravity waves this reduced Hamiltonian equations describe four-wave interaction.
In the case of statistical description of the wave field, Hasselmann kinetic equation~\cite{Hasselmann1962} for
the distribution of the wave action $n(k,t)=\langle|a_{\vec{k}}(t)|^2\rangle$ is used. Here
\begin{equation}
\label{a_k_def}
a_{\vec k} = \sqrt{\omega_k/(2k)} \eta_{\vec k} + \I \sqrt{k/(2\omega_k)} \psi_{\vec k},
\end{equation}
are complex normal variables. For gravity waves $\omega_k = \sqrt{gk}$.
More precisely, one has to use different function $b_{\vec k}$ after canonical transformation
eliminating nonresonant cubic terms~\cite{ZLF1992,Nazarenko2011} in the Hamiltonian, but the relative difference between corresponding $n_k$'s
in the case of $\mu\approx 0.1$ is of the order of few percents, so we shall limit ourself by this simpler function~\eqref{a_k_def}.

From the WTT~\cite{ZLF1992,Nazarenko2011}, in the case of four-waves interaction (which is the major interaction for surface gravity waves),
besides equipartion (Rayleigh-Jeans) spectrum, under few reasonable assumptions, one can find two constant flux Kolmogorov-Zakharov (KZ) solutions~\cite{ZF1967,ZakharovPhD,ZZ1982} of the Hasselmann kinetic equation:
\begin{equation}
\label{direct_inverse_cascade}
n_k^{(1)} = C_1 P^{1/3} k^{-\frac{2\beta}{3} - d},\;\; n_k^{(2)} = C_2 Q^{1/3} k^{-\frac{2\beta - \alpha}{3} - d}.
\end{equation}
For surface gravity waves, a coefficient of homogeneity of nonlinear interaction matrix element $\beta=3$,
the power of dispersion law $\alpha=1/2$, and the dimension of the surface $d=2$. As a result we get
\begin{equation}
\label{weak_turbulent_exponents}
n_k^{(1)} = C_1 P^{1/3} k^{-4},\;\;\; n_k^{(2)} = C_2 Q^{1/3} k^{-23/6}.
\end{equation}
The first solution $n_k^{(1)}$ describes direct cascade of energy from large pumping to small dissipative scales and was proven
to appear in simulations~\cite{Onorato2002,DKZ2003grav,DKZ2004,Korotkevich2008PRL}.
The second solution $n_k^{(2)}$ describes inverse cascade of wave action (or ``number'' of waves) from
small pumping to larger scales.

\section{Numerical scheme parameters}
We simulate equations~\eqref{eta_psi_system} in a (double) periodic box $L_x = L_y = 2\pi$. Grid resolution $N_x=N_y=512$.

Pumping on large scales (terms with $f_k$ in~\eqref{eta_psi_system}) and dissipation on small scales (terms with $\gamma_k$) is defined as follows:
\begin{equation}
\label{pumping_damping}
\begin{array}{lcl}
\displaystyle
f_k &=& 4 F_0\E^{\I R_{\vec k} (t)} \frac{(k-k_{p1})(k_{p2}-k)}{(k_{p2} - k_{p1})^2};\\
\displaystyle
D_{\vec k} &=& \gamma_k \psi_{\vec k},\;\;
\gamma_{k} = \begin{cases}
\gamma_0 (k - k_d)^2, 
k \ge k_d,\\
\gamma_{k} = 0, k < k_d.
\end{cases}
\end{array}
\end{equation}
Pumping parameters: $F_0 = 5\times10^{-9}(\times 2,\times 4,\times 8), k_{p1} = 60, k_{p2} = 64$. In other words, for four different simulation runs
the amplitude of a pumping function $|F_0|$ was differing by factors $\times 2$, $\times 4$, and $\times 8$ from the smallest one.
Function $f_k$ is a parabola with zeros
at $k_{p1}$ and $k_{p2}$ and magnitude extremum equal to $|F_0|$ in the middle $k_p=62$ between them; $f_k$ is zero outside of an interval $k\in[k_{p1};k_{p2}]$. $R_{\vec k} (t)$ -- uniformly distributed random number in interval $(0,2\pi]$, different for every harmonic and time step. Initial amplitudes for all $\eta_{\vec k}$ and $\psi_{\vec k}$ harmonics were
$10^{-12}$, all phases were uniformly distributed random numbers between $(0,2\pi]$.
Damping starts at $k_d=128$ and zero for larger scales (in order to avoid influence of aliasing due to cubic nonlinearity in our equations, we have to suppress harmonics with $k>k_{max}/2$,
where $k_{max}=256$). According to~\cite{DDZ2008} damping has to be included
in both equations of~\eqref{eta_psi_system}.
The value of $\gamma_0$ was chosen automatically to ensure 6 orders of harmonics magnitude difference between center of pumping region $k_p=62$
and the last Fourier harmonics at $k=k_{max}$. It guarantees good quality of the solution in spite of the fact that Fourier series being truncated. Later, after formation of the condensate, the difference of the largest and smallest (in absolute value) harmonics reaches 7 orders of magnitude. It should be noted, that values of $\gamma_0$ were different for different levels of pumping as different flux of energy to the high $k$'s had to be dissipated.
Details of the numerical algorithm can be found in~\cite{KDZ2016}.

In the $k$-space supports of $\gamma_{k}$ and $f_{k}$ are separated by the inertial
interval, where the Kolmogorov-type solution corresponding to direct cascade of energy could be recognized, but the range of scales in this particular set of numerical simulations was insufficient. Another inertial interval is located between $k=0$ and the pumping region, here we expected
to observe inverse cascade. Because both dissipation and pumping are isotropic (except for random phases) with respect to polar angle, we expect the same property for a solution. Even more, we use it for averaging the resulting spectra to replace ensemble averaging as it will be discussed . 

Computations were performed on a designated CPU core for each value of $F_0$ and took more than a year. This is the reason for
a relatively small grid resolution, even with respect to our previous computations~\cite{DKZ2004,Korotkevich2008PRL,Korotkevich2012MCS}.
During previous
works it had become clear that formation of the inverse cascade is an extremely slow process, which is explained by the fact that matrix
element of interaction for gravity waves behaves as $\sim k^3$, which means that interaction slows down really fast with decrease of $k$. Another fact is that in weakly nonlinear approximation characteristic nonlinear time $T_{nl}$, when nonlinearity shows itself (amplitude change of order 1), has to be much less
than a period corresponding to linear dispersion $T_k=2\pi/\omega_k$, and corresponding linear frequency $\omega_k$ also decays with decrease of $k$. We could not use consecutive doubling of resolution similarly to what we have done in our previous works, starting from~\cite{DKZ2003grav}, because in order
to avoid direct drain of energy from the pumping region through slave harmonics, we had to ensure that only third harmonic of pumping is
in dissipation region, otherwise the level of pumping and fluxes would strongly depend on the level of dissipation. Importance of such a mechanism was demonstrated
in~\cite{ZKPR2007,KPRZ2008}. At the same time we had to leave enough space for inverse cascade development. Taking into account all these considerations
and enormous time of the inverse cascade formation (in the end we had to compute till times $\sim 10^6 T_p$, where $T_p=2\pi/\omega_{k_p}$), the relatively small resolution $512\times 512$ was a reasonable compromise.

\section{Numerical results.}
\begin{figure}[ht!]
\centering
\includegraphics[width=3.5in]{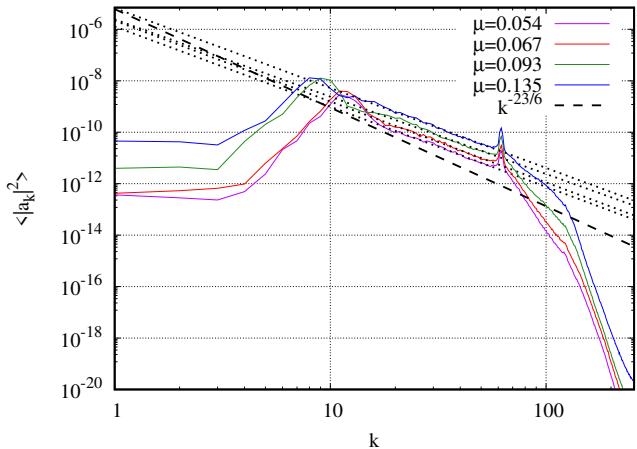}
\caption{\label{fig:All_angle_avrg} All simulations spectra (solid lines) with corresponding least squares fits (dotted lines, values of slopes are given in Table~\ref{tab:LSFit}) and KZ-spectrum slope (dashed line).}
\end{figure}

\begin{figure}[hb!]
\centering
\includegraphics[width=3.5in]{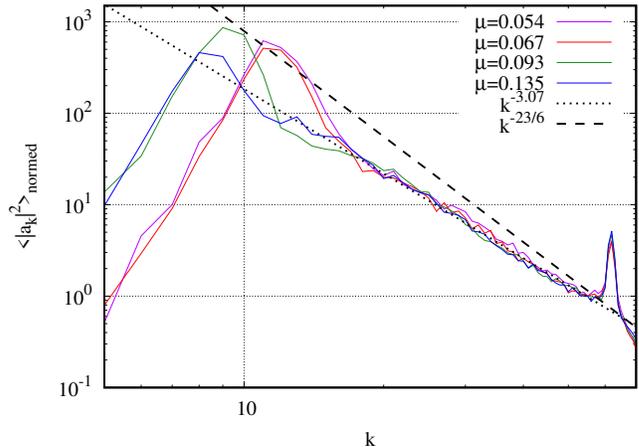}
\caption{\label{fig:All_angle_avrg_normed} All simulations spectra (solid lines) normed to have value of $\langle |a_k|^2\rangle$ at $k=55$ to be $1$ with a least squares fit (dotted line, value of the slope is given in Table~\ref{tab:LSFit}) over data points between condensate and pumping regions for all spectra and KZ-spectrum slope (dashed line).}
\end{figure}

The computations were performed till time corresponding to more than a million of a period of a harmonic at the maximum of pumping function $T_{p}$.
As a replacement of ensemble averaging, which is unfeasible taking into account computation time, in order to compute $\langle|a_{k}|^2\rangle$
we used averaging over an angle as the situation is isotropic and pumping has a random phase in every harmonics and at every moment of time.
Clearly, we have more harmonics to average over in larger $k$'s, meaning less fluctuations of averaged function. The resulting average steepnesses
$\mu$ for all four cases were: $0.054$, $0.067$, $0.093$, and $0.135$.
It has to be noted that these values correspond to a very different dissipation due to nonlinear processes, as was shown in~\cite{ZKP2009,KPZ2019}.
If we would have large enough dynamical range for development of a direct cascade, according to~\cite{Korotkevich2008PRL}, it would result in different slopes
for the direct cascade spectrum. The obtained angle averaged spectra are shown in Figure~\ref{fig:All_angle_avrg}.
One can see, that in accordance with the theory in~\cite{KDZ2016}, the width of the resonant curve, necessary for working quasi-resonances, depends on the nonlinearity in the system, resulting in further propagation of the inverse cascade and condensate for higher levels of steepness.   
For intervals of scales between regions influenced
by condensate and pumping, we observe power-like spectra (shown by dotted lines), with slopes practically independent of steepness. As one can see from the Figure~\ref{fig:All_angle_avrg}, the higher is nonlinearity level, the longer is inertial interval, where power-like spectrum can be observed.
This apparently universal spectrum has a slope which is even visually different from KZ-spectrum $\sim k^{-23/6}\approx k^{-3.83}$. The slopes' values together with intervals used for
linear least squares fit (in double logarithmic representation, like in Figure~\ref{fig:All_angle_avrg}) are given in Table~\ref{tab:LSFit}. The least squares fit was performed both in Octave~\cite{Octave} package (via standard $QR$-approach) and in Gnuplot~\cite{Gnuplot} (via Marquardt-Levenberg algorithm), standard errors~\cite{GnuplotFitErrors} are from Gnuplot computations.
\begin{table}[htb!]
\center
\begin{tabular}{|c|ccc|}
\hline
$\mu$ & $k\in$ & Average slope & Slope error \\
\hline
$0.054$ & $[17;55]$ & $-3.12$ & $\pm 0.04$ \\
\hline
$0.067$ & $[16;55]$ & $-3.14$ & $\pm 0.05$ \\
\hline
$0.093$ & $[12;56]$ & $-3.01$ & $\pm 0.05$ \\
\hline
$0.135$ & $[11;56]$ & $-3.11$ & $\pm 0.04$ \\
\hline
All & $170$ points & $-3.07$ & $\pm 0.02$ \\
\hline
\end{tabular}
\caption{\label{tab:LSFit} Least squares fits for different simulation spectra. The second column shows the range of $k$ between the condensate and pumping influence regions; the third column gives average slope $\alpha$ for $\langle |a_k|^2\rangle\sim k^{\alpha}$; the last column shows an estimated error of the fit.}
\end{table}
As one can see, these values are significantly different from $-23/6\approx -3.83$ in accordance with Figure~\ref{fig:All_angle_avrg}. In order to understand better the universality of the observed inverse cascade spectrum, we normalized all result in order
to make an amplitude of $|a_k|^2$ of the harmonic $k=55$ equal to 1. This is the harmonic closest to the region influenced by pumping which is present in all
datasets for least squares fits in Table~\ref{tab:LSFit}. As a result, all spectra collapsed to a single curve (line) in the inertial interval (up to variability
due to insufficient averaging), as can be seen in Figure~\ref{fig:All_angle_avrg_normed}. We used the set of all available points in inertial intervals
for all simulation runs for a least squares fit. The resulting slope is given in the last line of the Table~\ref{tab:LSFit} and shown as a dotted line
in Figure~\ref{fig:All_angle_avrg_normed}.
As one can see, the value of the estimated  spectrum slope is $\langle |a_k|^2\rangle\sim k^{-3.07}$, which is again, expectedly, different from
the one predicted by the WTT(see~\eqref{weak_turbulent_exponents}). The reason for this inconsistency is yet to be found. One of the factors which might be important is the deviation of
the dispersion relation from the linear one in the presence of the condensate (according to~\eqref{direct_inverse_cascade}, the slope of the inverse cascade spectrum directly depends upon $\alpha$, which is the power of dispersion relation). It was show in~\cite{Korotkevich2013JETPL} that
in the presence of condensate rotation of phase of harmonics in the inverse cascade region cannot be described with a reasonable accuracy only by
linear dispersion relation for gravity waves $\omega_k=\sqrt{gk}$. One can try to take into account interaction with condensate, e.g. using Bogolyubov transformation like in the case
of dilute Bose gas (see sections about degenerated almost ideal Bose gas in~\cite{AGD1962} or~\cite{LL_vol9}).

It also can be noted that Figure~\ref{fig:All_angle_avrg_normed} strongly resembles Figure~3 in~\cite{AS2006PRL}, with the difference that we performed four different simulations and have a significantly longer inertial interval for inverse cascade which allows us to determine the slope with a reasonable accuracy. Values in Table~\ref{tab:LSFit} with given errors have one point which is common for all but one simulation runs (documentation for Gnuplot {\tt fit} function recommends to consider
standard errors just as an estimation, as many assumptions were used which cannot be justified in our case
), corresponding to the slope $-3.09$. Thus, based on our results, one can conclude,
that the slope is roughly between $-3.0$ and $-3.1$, most probably closer to the $-3.1$ value. Nevertheless, difference between observed slope and the one predicted by the WTT is clear both visually and numerically.

\section{Conclusion}
During four numerical simulations with different pumping and dissipation parameters, we observed formation of the inverse cascade accompanied by the condensate. In the inertial interval between condensate and pumping regions the spectrum is close to a power-like function with a least square fit suggesting $\sim k^{-3.07}$ slope. This result is close to the experimentally observed spectrum in recent wave tank experiments~\cite{NL2016} (see Figure~12 (right) where our $n_k\sim k^{-3.07}$ would correspond to $E_k\sim k^{-1.57}$ which is close to the $E_k\sim k^{-1.5}$ proposed in that paper), especially taking into account relatively noisy experimental data and short range of scales. The slope of the spectrum is virtually identical for dramatically different levels of nonlinearity, which suggests universality of the observed solution. Because spectrum is significantly different from the one predicted by the WTT ($\sim k^{-23/6}\approx k^{-3.83}$), the dependence of a constant in front of the power-like function on the pumping parameters was not investigated.
The applicability of WKE (which is derived for an infinite domain) to the DNS in a periodic box is yet to be investigated in details. Recent works on 3D nonlinear Schr\"{o}dinger equation (NLSE)~\cite{ZSKN2022} and 1D quintic NLSE~\cite{BBKKS2022} give us hope that precise range of simulation parameters, when one could expect quantitative correspondence between WKE and DNS of dynamical equations for surface gravity waves, will be determined in future works. For now we could use previous simulations~\cite{Onorato2002,DKZ2004,AS2006JFM,ZKPR2007,Korotkevich2008PRL} as an empirical evidence that we could expect at least qualitative correspondence (spectra slopes, spectrum peak downshift etc.) for simulations in the frameworks of these significantly different models.

It has to be mentioned that WTT constant wave action flux KZ-spectrum $\sim k^{-23/6}$ was derived for a very specific case (infinite inertial interval, one spectrum in the whole range of wavenumbers, etc.), which is different from what we observe in our simulations (finite range of scales, limited both by condensate and pumping). Similarly, the direct cascade of energy KZ-spectrum was not always observed in wave tank experiments~\cite{Lukaschuk2007,LNMD2009}, while in the open water in most of the cases the observed spectrum corresponds to KZ-spectrum from WTT. The explanation of the spectrum reported in this Letter is yet to be proposed, although previous works~\cite{Korotkevich2008PRL,Korotkevich2012MCS,Korotkevich2013JETPL} give us a hint that presence of the condensate plays a major role for processes in the inverse cascade inertial interval. In order to take into account the condensate, one could try to use an approach (Bogolyubov transformation) similar to one used in a recent paper~\cite{GKLN2022}. At the same time everything depends on the type of the condensate. If it is a coherent structure, Bogolyubov transformation is a method of choice. Because we have the condensate as a ring with a radius between $10$ and $15$ for different simulations and width around few harmonics, total number of discreet harmonics in condensate is of the order of a hundred and if they are stochastic enough (this is yet to be defined) the situation could be described by WKE. In both of these cases complexity of the interaction coefficient (e.g. see Appendix~B in~\cite{PRZ2003}) for surface gravity waves makes analysis difficult enough to be a topic of a separate investigation.

\section{Acknowledgements}
The author is grateful for support from the Simons' Collaboration on Wave Turbulence (award \#651459). The simulations presented in this article were performed using the Landau Institute for Theoretical Physics computational resources. The paper was written during author's visit to the the Universit\'e C\^ote d'Azur/Institut de Physique de Nice, France, funded by F\'ed\'eration de Recherche ``Wolfgang D\"oblin'' and ``Waves Complexity'' visiting researcher program, to whom author is thankful for hospitality and support. The author would also like to thank V.\,V.~Lebedev and I.\,V.~Kolokolov for stimulating discussions. This work would not be possible without free software from GNU Project~\cite{GNU}, Gnuplot~\cite{Gnuplot}, and FFTW library~\cite{FFTW}.

\appendix

\section{Supplemental Material: Exact equations.}
\label{intro}
\subsection{Laplace equation formulation.}
Let us consider three dimensional irrotational flow of an inviscid (ideal) incompressible
and homogeneous fluid of infinite depth. Because fluid is irrotational $\curl\vec v = \vec 0$,
we can introduce velocity potential $\Phi=\Phi(x,y,z; t)$ in the following way: $\vec v = \vec \nabla\Phi$.
Because fluid is incompressible $\diver \vec v = 0$ and
velocity potential $\Phi$ satisfies the Laplace equation
\begin{equation}
\label{Laplace}
\diver \vec v = \Delta \Phi = 0
\end{equation}
in the domain filled by fluid
\begin{equation}
-\infty < z < \eta (\vec r), \;\;\;\; \vec r = (x,y).
\end{equation}
Here and further $\Delta = \vec \nabla^2$ and $\eta = \eta (x,y,t)$ is a fluid surface elevation
with respect to a steady state (flat horizontal surface positioned at $z=0$).

Boundary conditions for velocity potential are as follows
\begin{equation}
\label{Laplace_boundary}
\begin{array}{c}
\displaystyle
\frac {\pD \eta}{\pD t}
= \left.\left(\frac{\pD \Phi}{\pD z} -
\frac {\pD \Phi}{\pD x}\frac {\pD \eta}{\pD x} -
\frac {\pD \Phi}{\pD y}\frac {\pD \eta}{\pD y}\right) \right|_{z= \eta},\\
\displaystyle
\left. \left ( 
\frac {\pD \Phi}{\pD t}+
\frac{1}{2}( \vec \nabla \Phi )^2
\right ) \right |_{z= \eta} + p|_{z=\eta} + \rho g\eta = 0,\\
\displaystyle
p|_{z=\eta} = \sigma \vec \nabla \cdot \frac{\vec \nabla \eta}{\sqrt{1+(\vec \nabla \eta)^2}}
\end{array}
\end{equation}
\begin{equation}
\label{Laplace_boundary_bottom}
\phi_z|_{z\rightarrow - \infty} = 0.
\end{equation}
Also it is usually reasonable to suppose that all velocities at infinities are zeros as well as surface elevation. We will discuss periodic box case later.
Here we introduced gravity acceleration $g$, fluid density $\rho$, and surface tension coefficient $\sigma$.

\subsection{Hamiltonian equations formulation.}
Kinetic and potential energies of the system are the following
$$
H = T + U,
$$
\begin{equation}
\label{kinetic_full}
T = \frac{\rho}{2} \int \D^2 r \int \limits_{-\infty}^{\eta} (\vec \nabla \Phi)^2 \D z,
\end{equation}
\begin{equation}
\label{potential_full}
U = \rho\int \left(\frac{\sigma}{\rho}\left(\sqrt{1 + (\vec \nabla \eta)^2} - 1\right) + \frac{g}{2} \eta^2\right) \D^2 r.
\end{equation}
Let us introduce $\sigma' = \sigma/\rho$, then one can get rid of $\rho$ after renormalization of energies. It will
correspond to rescaled time $t\rightarrow \rho t$.

System (\ref{Laplace_boundary}), (\ref{Laplace_boundary_bottom}) has a Hamiltonian
structure. It was shown by Zakharov in 1968~\cite{Zakharov1968} that using variables $\eta(x,y,t)$-surface
displacement and velocity potential on the surface $\psi (x,y;t) = \Phi (x,y,\eta(x,y;t);t)$
these boundary conditions take form
\begin{equation}
\label{Hamiltonian_equations}
\frac{\pD \eta}{\pD t} = \frac{\delta H}{\delta \psi}, \;\;
\frac{\pD \psi}{\pD t} = - \frac{\delta H}{\delta \eta}.
\end{equation}
Thus variables $\eta, \psi$ are canonically conjugated.

Kinetic energy cannot be expressed in terms of $\eta, \psi$ in an explicit form. However,
one can find expansion of the Hamiltonian $H$ in powers of nonlinearity. Let use the following vector identity:
$$
\vec\nabla \Phi \cdot \vec \nabla\Phi = \diver(\Phi\vec\nabla\Phi) - \Phi\Delta\Phi = \diver(\Phi\vec\nabla\Phi).
$$
Thus, using Stokes' theorem kinetic energy~\eqref{kinetic_full} can be rewritten as an integral over the surface:
\begin{equation}
2 T = \int (\vec\nabla\Phi)^2 \D V = \int \diver(\Phi\vec\nabla\Phi) \D V = \int \Phi\vec\nabla\Phi \cdot \D\vec S.
\end{equation}
Here we have to take into account the fact that on infinitely remote boundaries all velocities has to be zeros,
thus $\vec\nabla\Phi \equiv \vec 0$ everywhere but on the upper fluid surface. If we consider periodic boundary conditions for $x$ and $y$, then outgoing flux will be exactly compensated by incoming due to periodicity. So this surface integral has to be understood
as an integral over the (upper) surface of the fluid. On the surface radius vector is $\vec R = x\vec i + y\vec j + \eta(x,y)\vec k$, where $\vec i$, $\vec j$, and $\vec k$ are unit vectors (orts) along directions of $x$-, $y$-, and $z$-axes correspondingly.
and for the oriented surface element one can write (we follow an agreement that normal vector is directed outside of the enclosed volume, which for upper surface means that $z$-component of the normal vector is mostly positive):
$$
\D \vec S = \frac{\pD\vec R}{\pD x}\times\frac{\pD\vec R}{\pD y} \D x \D y=
\left(\vec k - \vec i \frac{\pD\eta}{\pD x} - \vec j \frac{\pD\eta}{\pD y}\right)\D x \D y.
$$
Using this expression of $\D \vec S$ one can write
\begin{align}
\label{kinetic_nabla_phi_nabla_eta}
2 T &= \int \Phi\vec\nabla\Phi \cdot \D\vec S\nonumber\\
&=\int\Phi
\left.\left(\frac{\pD\Phi}{\pD z} - \frac{\pD\Phi}{\pD x}\frac{\pD\eta}{\pD x} - \frac{\pD\Phi}{\pD y} \frac{\pD\eta}{\pD y}
\right)\right|_{z=\eta}\D x \D y\nonumber\\
 &= \int\Phi
\left.\left(\frac{\pD\Phi}{\pD z} - \vec\nabla\Phi\cdot\vec\nabla\eta
\right)\right|_{z=\eta}\D x \D y
\end{align}
Pay attention that expression in the brackets is normal velocity $v_{\vec n}$ with some factor:
$$
\left.\left(\frac{\pD\Phi}{\pD z} - \frac{\pD\Phi}{\pD x}\frac{\pD\eta}{\pD x} - \frac{\pD\Phi}{\pD y} \frac{\pD\eta}{\pD y}
\right)\right|_{z=\eta} = v_{\vec n}[1+(\vec\nabla\eta)^2],
$$
and gives insight in the first (kinematic) boundary condition~\eqref{Laplace_boundary}. You can see that this boundary condition requires that the surface has to move with the same velocity as the fluid.
 
As a first step let us take care of the second term. Using the definition of potential on the surface $\psi$
one can get the following relations
\begin{align*}
\frac{\pD \psi}{\pD x} = \frac{\pD}{\pD x} \Phi(x,y,\eta(x,y)) = \left.\frac{\pD\Phi}{\pD x}\right|_{z=\eta} + \left.\frac{\pD\Phi}{\pD z}\right|_{z=\eta}\frac{\pD\eta}{\pD x},\\
\frac{\pD \psi}{\pD y} = \frac{\pD}{\pD y} \Phi(x,y,\eta(x,y)) = \left.\frac{\pD\Phi}{\pD y}\right|_{z=\eta} + \left.\frac{\pD\Phi}{\pD z}\right|_{z=\eta}\frac{\pD\eta}{\pD y}.
\end{align*}
Using these relations we immediately get
\begin{align}
\left.\left(\frac{\pD\Phi}{\pD x}\frac{\pD\eta}{\pD x} + \frac{\pD\Phi}{\pD y} \frac{\pD\eta}{\pD y}
\right)\right|_{z=\eta} \nonumber\\
= \vec\nabla\psi\cdot\vec\nabla\eta - \left.\frac{\pD\Phi}{\pD z}\right|_{z=\eta}(\vec\nabla\eta)^2,\label{nabla_phi_nabla_eta}
\end{align}
which after substitution of \eqref{nabla_phi_nabla_eta} into \eqref{kinetic_nabla_phi_nabla_eta} yields
\begin{align}
&T = \frac{1}{2}\int (\vec\nabla\Phi)^2 \D V \nonumber\\
&=\frac{1}{2}\int\Phi
\left.\left(\frac{\pD\Phi}{\pD z}[1+(\vec\nabla\eta)^2] - \vec\nabla\psi\cdot\vec\nabla\eta
\right)\right|_{z=\eta}\D x \D y.\label{kinetic_surface_exact}
\end{align}
Let us emphasize the fact that up to now all derivations are exact, we haven't used the weak nonlinearity in the system yet.

\section{Supplemental Material: Expansion of kinetic energy in terms of steepness.}
\subsection{Steepness as a measure of nonlinearity in the system.}
From observation it is known that in most of the interesting cases steepness (average slope) of the surface $\mu$
is of the order of the value 0.1 or lower. Average slope can be introduced in numerous ways. Here are just few of them:
\begin{itemize}
\item $\mu = \sqrt{\langle|\vec\nabla\eta|^2\rangle}$.
\item $\mu = \langle|\vec\nabla\eta|\rangle$.
\item $\mu = \sqrt{\langle\eta^2\rangle} k_p$.
\end{itemize}
Here $k_p = 2\pi/\lambda_p$ is wavenumber corresponding to the characteristic wavelength of the wave field $\lambda_p$.
In terms of wavenumber Fourier spectrum of the wave field it is wavenumber of the peak of the spectrum.
Because we shall be working with the weakly nonlinear wave field it is natural to use Fourier transform in $xy$-plane. Let's introduce it
for function of two variable $f=f(x,y)=f_{\vec r}$:
\begin{align}
\hat F [f_{\vec r}] &= f_{\vec k} = \frac{1}{(2\pi)^2} \int f_{\vec r} e^{\I {\vec k} {\vec r}} \D^2 r,\label{Forward_Fourier}\\
\hat F^{-1} [f_{\vec k}] &= f_{\vec r} = \int f_{\vec k} e^{-\I {\vec k} {\vec r}} \D^2 k.\label{Inverse_Fourier}
\end{align}
For periodic boundary conditions, obviously, these Fourier integrals should be replaced by Fourier series. In the case of $2\pi\times 2\pi$
periodic box Fourier series and these Fourier integrals will coincide.
Now let us try to get expansion of kinetic energy up to the quartic terms.

From the Laplace equation~\eqref{Laplace} after Fourier transform in $xy$-plane one gets equation
$$
\frac{\pD^2 \Phi_{\vec k}}{\pD z^2} - k^2 \Phi_{\vec k} = 0,
$$
which together with boundary conditions at infinite depth (solution has to decay for negative $z$) immediately yields
\begin{equation}
\label{Phi_Phi_z_k}
\Phi_{\vec k} = A_{\vec k}\E^{kz},\;\; \frac{\pD\Phi_{\vec k}}{\pD z} = kA_{\vec k}\E^{kz}=k\Phi_{\vec k}.
\end{equation}
Here we introduced magnitude of the wavevector $k=|\vec k|$.

\subsection{Expansion of potential at the surface.}
Now let us express $A_{\vec k}$ through the functions on the surface. In order to do this we expand exponential function
in~\eqref{Phi_Phi_z_k} up to the second order terms:
\begin{equation}
\label{Phi_k_exp}
\Phi_{\vec k} \simeq A_{\vec k} \left(1 + kz + \frac{1}{2}(kz)^2\right).
\end{equation}
Pay attention, that on the surface $z=\eta$ and $kz$ is of the order of steepness $\mu$, which is our small parameter.
We limit ourselves by quadratic terms, because together with amplitude $A$ they will give cubic term for $\Phi$,
while in the kinetic energy~\eqref{kinetic_surface_exact} derivative $\pD \Phi/\pD z$,
which will be of the same order as $\Phi$,
is multiplied by potential $\Phi$, which will result in quartic terms in Hamiltonian. Further expansion will
give us higher order terms.

After inverse Fourier transform of~\eqref{Phi_k_exp} expanded potential takes the form
\begin{align}
\Phi (x,y,z) &= \hat F^{-1}[\Phi_{\vec k}] \simeq A(x,y) \nonumber\\
&+z\hat F^{-1}[kA_{\vec k}] + \frac{z^2}{2}F^{-1}[k^2A_{\vec k}],\nonumber
\end{align}
which on the surface gives
\begin{align}
\psi(x,y) &= \Phi (x,y,z)|_{z=\eta} = A(x,y) \nonumber\\
&+ \eta\hat F^{-1}[kA_{\vec k}] + \frac{\eta^2}{2}F^{-1}[k^2A_{\vec k}],
\end{align}
here $A(x,y) = \hat F^{-1}[A_{\vec k}]$ and from now on we shall replace $\simeq$ sign with equality.

It is convenient to introduce linear nonlocal operator $\hat k$ in the following way:
\begin{equation}
\label{k_operator}
\hat k f(x,y) = \hat F^{-1}[k\hat F[f_{\vec r}]] = \hat F^{-1}[ k f_{\vec k}].
\end{equation}
In other words, this operator acts as follows:
it multiplies Fourier harmonics by the magnitude of corresponding wave number. Due to this property it is often
referred as a square root of negative Laplacian: $\hat k = \sqrt{-\Delta}$. With this operator we get the following compact
notation:
\begin{equation}
\psi(x,y) = A + \eta \hat k A  - \frac{\eta^2}{2} \Delta A,
\end{equation}
In order to calculate $\pD \Phi/\pD z$ on the surface, according to~\eqref{Phi_Phi_z_k} we need to find $A$.
Let us write equation for $A$:
\begin{equation}
\label{A_equation}
A = \psi - \eta \hat k A  + \frac{\eta^2}{2} \Delta A,
\end{equation}
this is integro-differential nonlocal equation. Let us solve it by iterations, keeping terms up to cubic ones.
In the first iteration we put $A=\psi$ in the right hand side of~\eqref{A_equation}:
\begin{equation}
\label{A_first_iter}
A = \psi - \eta \hat k \psi  + \frac{\eta^2}{2} \Delta \psi.
\end{equation}
In the second iteration we substitute this result~\eqref{A_first_iter} in the right hand side of~\eqref{A_equation} (omit terms higher than cubic ones):
\begin{equation}
\label{A_second_iter}
A = \psi - \eta \hat k \psi  + \frac{\eta^2}{2} \Delta \psi + \eta \hat k [\eta\hat k\psi].
\end{equation}
This is our solution with desired accuracy. Now we can use it for $\pD \Phi/\pD z$ on the surface. 

\subsection{Expansion of $\pD \Phi/\pD z$ at the surface.}
Let us apply exactly the same approach to the $z$-derivative of potential. From~\eqref{Phi_Phi_z_k} and~\eqref{Phi_k_exp} one gets (up to the cubic terms):
\begin{equation}
\label{Phi_z_k_exp}
\frac{\pD\Phi_{\vec k}}{\pD z}(x,y,z) \simeq k A_{\vec k} \left(1 + kz + \frac{1}{2}(kz)^2\right).
\end{equation}
After the inverse Fourier transform:
$$
\frac{\pD\hat F^{-1}[\Phi_{\vec k}]}{\pD z} \simeq \hat F^{-1}[kA_{\vec k}] + z\hat F^{-1}[k^2A_{\vec k}] + \frac{z^2}{2}F^{-1}[k^3 A_{\vec k}],
$$
which on the surface gives
\begin{equation*}
\left.\frac{\pD\Phi}{\pD z}\right|_{z=\eta} = \hat F^{-1}[kA_{\vec k}] + \eta\hat F^{-1}[k^2 A_{\vec k}] + \frac{\eta^2}{2}F^{-1}[k^3 A_{\vec k}].
\end{equation*}
This expression can be rewritten using $\hat k$-operator:
\begin{equation}
\left.\frac{\pD\Phi}{\pD z}\right|_{z=\eta} = \hat k A - \eta \Delta A - \frac{\eta^2}{2} \Delta[ \hat k A].
\end{equation}
Let us substitute here $A$ from expression~\eqref{A_second_iter} and keep only terms up to cubic ones:
\begin{align}
\left.\frac{\pD\Phi}{\pD z}\right|_{z=\eta} &= \hat k \psi - \eta\Delta \psi - \frac{1}{2}\eta^2\Delta\hat k\psi - \hat k [\eta \hat k \psi]\nonumber\\
&+ \eta\Delta(\eta\hat k \psi)
+ \frac{1}{2}\hat k[\eta^2 \Delta \psi] + \hat k[\eta \hat k [\eta\hat k\psi]].\label{Phi_z}
\end{align}
Now we have everything for rewriting Hamiltonian in terms of $\eta$ and $\psi$.

\section{Supplemental Material: Weakly nonlinear Hamiltonian equations.}
For given boundary conditions the following relation holds:
\begin{equation}
\label{nabla_by_parts}
\int\eta\vec\nabla\psi \D^2 r = - \int\psi\vec\nabla\eta \D^2 r.
\end{equation}
For operator $\hat k$ we can derive similar relation:
\begin{equation}
\label{k_by_parts}
\int\psi\hat k\eta \D^2 r = \int k \eta_{\vec k}\psi_{\vec k}^*\D^2 k  = \int (\hat k\psi) \eta \D^2 r,
\end{equation}
where we used the relation for Fourier image of a real valued function $\psi_{-\vec k} = \psi_{\vec k}^*$, while $*$ means complex conjugation.

Using~\eqref{nabla_by_parts}-\eqref{k_by_parts} and ~\eqref{Phi_z} kinetic energy~\eqref{kinetic_surface_exact} can be written in a relatively compact form:
\begin{align}
T = \frac{1}{2}\int\psi \hat k  \psi \D^2 r +
\frac{1}{2}\int\eta\left[ |\nabla \psi|^2 - (\hat k \psi)^2 \right] \D^2 r\nonumber\\
\label{Kinetic_expanded}
+ \frac{1}{2}\int\eta (\hat k \psi) \left[ \hat k (\eta (\hat k \psi)) + \eta\nabla^2\psi \right] \D^2 r.
\end{align}

Because we expand Hamiltonian up to quartic terms using small parameter $\mu \sim |\vec \nabla \eta | \sim |\hat k \eta|$ one can expand part of potential
energy associated with capillary waves:
\begin{equation}
\label{cap_expanded}
\sigma'\left(\sqrt{1 + (\vec \nabla \eta)^2} - 1\right) \simeq \sigma'\left(\frac{1}{2}(\vec \nabla \eta)^2 - \frac{1}{8}(\vec \nabla \eta)^4\right).
\end{equation}

Hamiltonian, resulting from~\eqref{potential_full}, \eqref{cap_expanded}, and~\eqref{Kinetic_expanded} have the following form
\begin{flalign}
H &= \frac{1}{2}\int\left(\sigma' |\nabla \eta|^2 + g \eta^2 + \psi \hat k \psi \right) \D^2 r \nonumber\\
&+\frac{1}{2}\int\eta\left[ |\nabla \psi|^2 - (\hat k \psi)^2 \right] \D^2 r \nonumber\\
&+ \frac{1}{2}\int\eta (\hat k \psi) \left[ \hat k (\eta (\hat k \psi))
+ \eta\nabla^2\psi \right] \D^2 r\nonumber\\
&+ \frac{1}{8}\int\sigma'(\vec\nabla\eta)^4\D^2 r.\label{Hamiltonian}
\end{flalign}

Hamiltonian equations (\ref{Hamiltonian_equations}) can be written as follows
\begin{flalign}
\dot \eta =& \hat k  \psi - (\nabla (\eta \nabla \psi)) - \hat k  [\eta \hat k  \psi] +
\hat k (\eta \hat k  [\eta \hat k  \psi]) \nonumber\\
&+ \frac{1}{2} \nabla^2 [\eta^2 \hat k \psi] + \frac{1}{2} \hat k [\eta^2 \nabla^2\psi],\nonumber\\
\dot \psi =& \sigma' \nabla^2 \eta - g\eta - \frac{1}{2}\left[ (\nabla \psi)^2 - (\hat k \psi)^2 \right]\nonumber\\
&-[\hat k  \psi] \hat k  [\eta \hat k  \psi]
- [\eta \hat k  \psi]\nabla^2\psi - \sigma'\diver((\vec\nabla\eta)^2 \vec\nabla\eta).
\end{flalign}
In many cases, when capillary waves are considered it is enough to limit ourselves by cubic terms in the Hamiltonian,
this is why the quartic capillary term is often dropped even when capillary effects are taken into account.

%
\end{document}